\documentclass{kluwer}
\usepackage[dvips]{graphicx}

\begin{document}
\newcommand \msun {\mbox{$\mathcal{M}_{\odot}$}}
\newcommand \kms {\mbox{km~s$^{-1}$}}
\newcommand \degree {\mbox{$^\circ$}}
\newcommand \micron  {\mbox{$\mu$m}}
\newcommand \solar {L$_{\odot}$\ }
\newcommand \solm {M$_{\odot}$}
\begin{article}
\begin{opening}
\title{The Warped Gas and Dust Lane in NGC~3718}
\subtitle{NGC~3718}

\author{M. \surname{Krips}\email{Krips@ph1.uni-koeln.de}}
\author{J.-U. \surname{Pott}}
\author{A. \surname{Eckart}}
\author{S. \surname{Leon}}
\author{C. \surname{Straubmeier}}
\institute{I. Physikalisches Institut Universtity of Cologne}




\runningtitle{NGC~3718}
\runningauthor{M. Krips}

\begin{ao}
Kluwer Prepress Department\\
P.O. Box 990\\
3300 AZ Dordrecht\\
The Netherlands
\end{ao} 


\begin{abstract} 
We present the first observations of molecular line emission
in $NGC~3718$ with the IRAM 30m and the Plateau de Bure Interferometer.
This galaxy is an impressive example for a strongly warped gas disk 
harboring an active galactic nucleus (AGN).
An impressive dust lane is crossing the nucleus and  a warp is developing
into a polar ring.
The molecular gas content is found to be typical of an elliptical galaxy with
a relatively low molecular gas mass content ($\sim 4\times10^8 \msun$).
The molecular gas distribution is found to warp
from the inner disk together with the HI distribution. The
CO data were also used to improve the kinematic modeling in the inner
part of the galaxy, based on the so-called {\em tilted ring}-model. The nature
of $NGC~3718$ is compared with its northern sky ``twin'' Centaurus A and
the possible  recent swallowing of a small-size gas-rich spiral is discussed.
\end{abstract}

\keywords{galaxies: individual: NGC~3718 -- galaxies: active --
galaxies: kinematics and dynamics -- ISM: molecules} 

%
%
\end{opening}

NGC~3718 is a peculiar galaxy at a distance of 13~Mpc.
The most prominent features of this galaxy are its large,
warped dust lane running across the entire stellar bulge of the galaxy.
The galaxy NGC~3718 (Arp 214) and its companion
NGC~3729 form a galaxy pair,
at a distance of about 13~Mpc\footnote{Throughout this
work we assume H$_0$=75~km~s$^{-1}$~Mpc$^{-1}$} as
part of the Ursa Major Galaxy Cluster.
In addition NGC~3718 and NGC~3729 were classified
as members of Group 241 in the LGG (Lyon group of galaxies catalog). Both the
cluster and the group are poorly defined, with velocity dispersions of less
than 150~km~s$^{-1}$. Such a low velocity dispersion is an indication, that
cannibalism of small galaxies is quite likely. 

The galaxies are also contained in the Hubble Atlas of Galaxies
and the Atlas of Peculiar Galaxies, where the extremely dark dust 
feature and a relatively small nucleus of NGC~3718 become 
apparent.
NGC~3718 shows its very peculiar optical appearance, with the
prominent dust lane, covering a large part of the galaxy's main body. 
It starts with a width of less than 2~arcsec at the center and extends 
into several smooth filaments across the bulge of the galaxy. 
At a separation of about 1.5~arcmin from the nucleus they bend by 
almost 90 degrees  towards the north and the south and remain 
visible over a distance of more than 6~arcmins (24~kpc).
Inspection of the dust lane shows that the warp signature goes all the way
into the center of the galaxy (Fig.1).
The warp is also present in the atomic gas and has been kinematically modeled
by Schwarz (1985).

\begin{figure}
\centerline{\includegraphics[width=15pc]{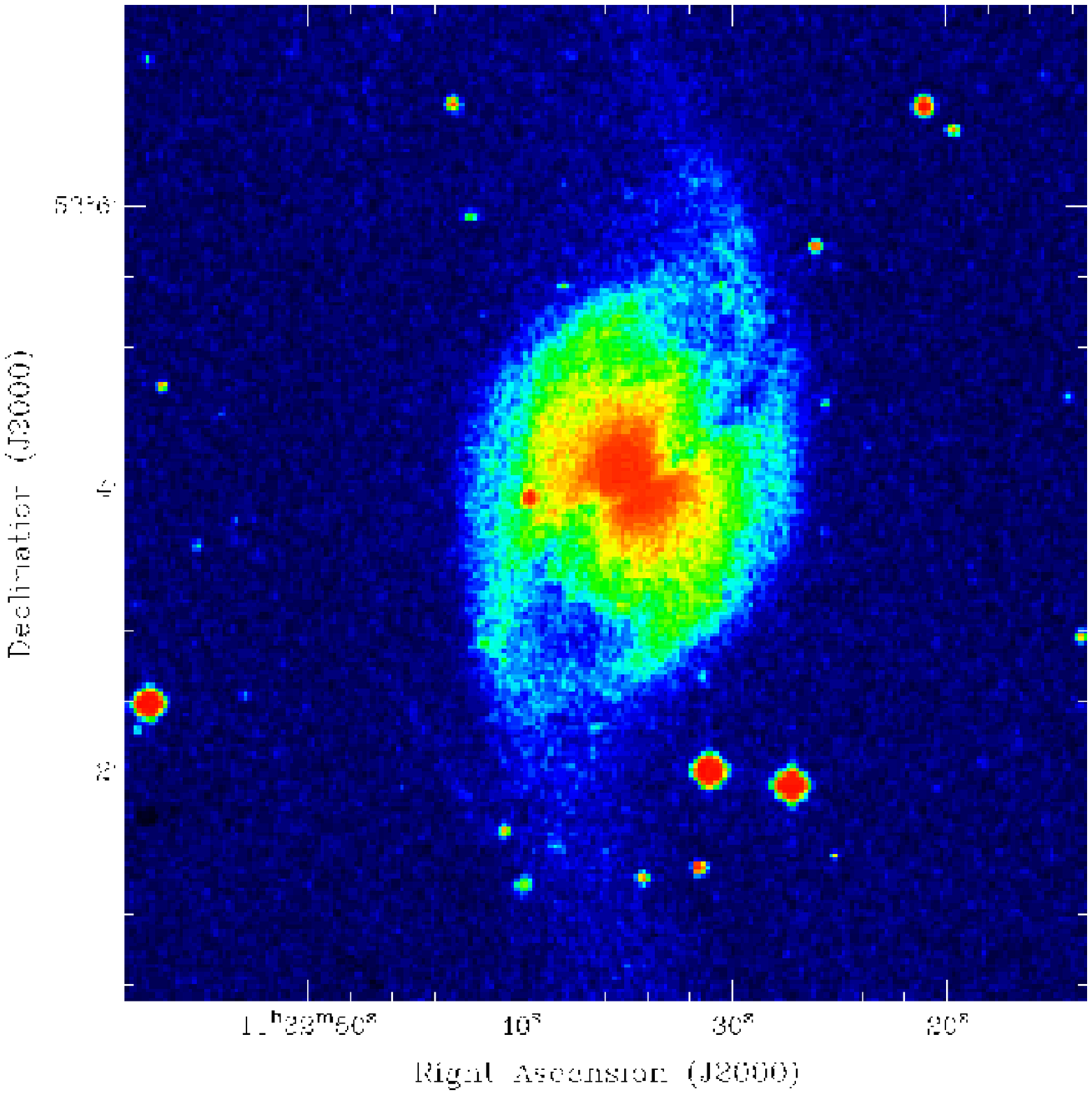} \includegraphics[width=15pc]{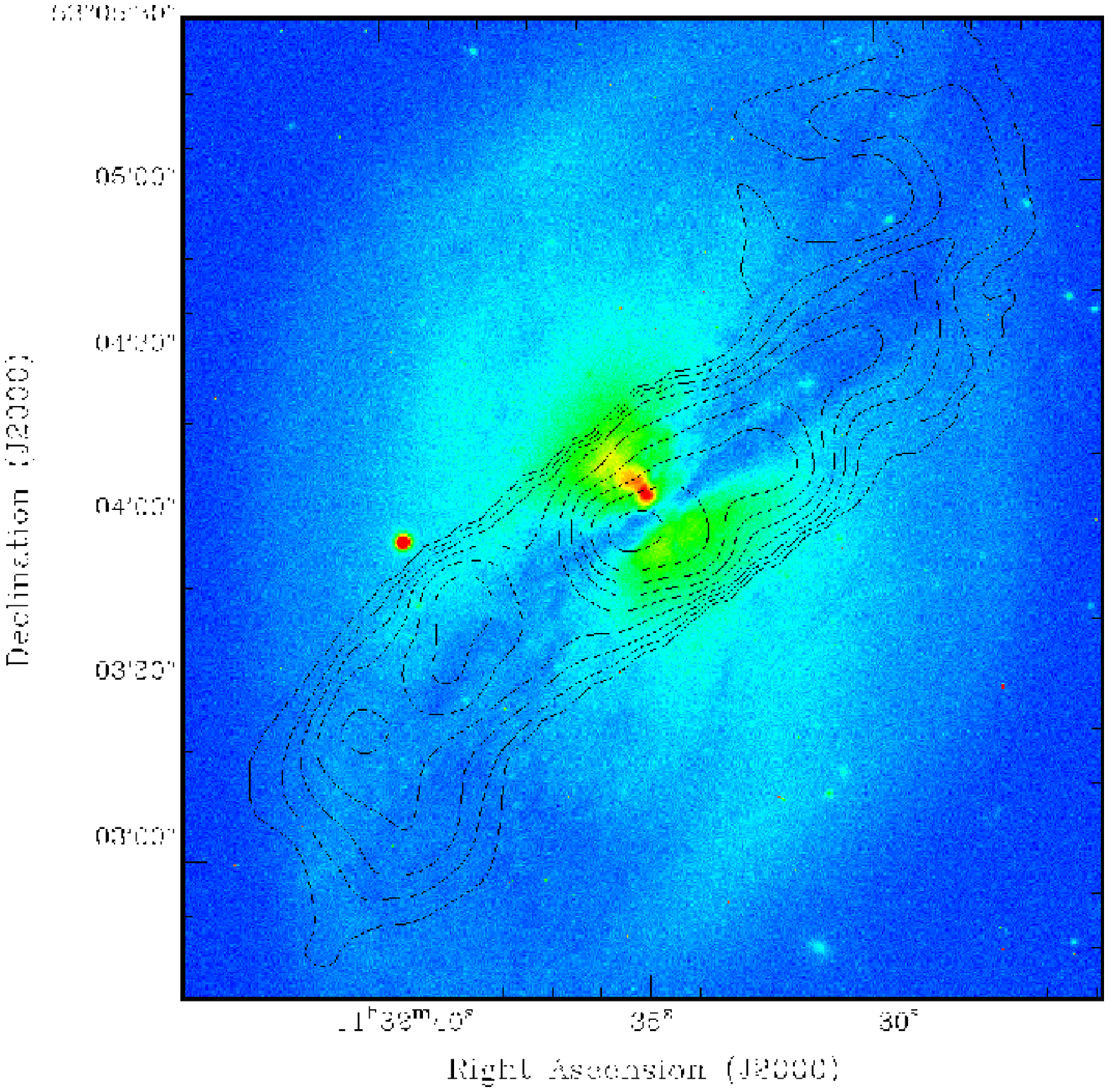}}
\caption{left panel:Optical image of NGC~3718. Taken from the DSS
  survey. right panel: The opical image, overlaid with the contours (linear scale-steps
of $\sigma\sim0.3\,K\,\frac{km}{s}$) of our CO(1-0) obersations. The warp
of the molecular gas distribution is as obvious as the coincidence of gas and dust lane.}
\end{figure}

NGC~3718 contains an active nucleus.
Ho et al. (1997) classified NGC~3718 as both a LINER and a
Seyfert 1.9 galaxy. The nucleus shows weak, broad H$\alpha$ 
emission with FWHM$=2350$~km~s$^{-1}$. Filippenko et al. (1985)
report strong [O~I]~$\lambda$~6300 with FWHM$=570$~km~s$^{-1}$ 
indicative for a hidden AGN (Ho et al. 1997).
Burke \& Miley (1973) found a radio source at the position of the nucleus
of NGC~3718, reporting a 1415~MHz flux density of $(20\pm 5) \cdot
10^{-29}$ Wm$^{-2}$Hz$^{-1}$. The radio emission is at the lower end of the
range of Seyfert galaxies, but exceeds that of most normal galaxies.
Using the HST, Barth et al. (1998) searched for ultraviolet emission
in LINER galaxies and did not detect NGC~3718. A possible explanation
could be that the UV sources are obscured by dust, 
which is reasonable in the case of NGC~3718.
From its optical appearance and the large scale dynamics
NGC 3718 can be regarded as the northern hemisphere counterpart of Centaurus~A.
In Centaurus~A also a warping of the molecular gas disk in the inner 1~kpc is
invoked to explain its dust lane and the associated complex kinematics
(Sparke 1996, Quillen, Graham, and Frogel 1993, and Quillen et al. 1992).
\\
The HI maps (Schwarz 1985) show a depletion of HI gas towards the central
arcminute right on the dust lane.
This is probably the location were - like in Cen~A
(Eckart et al. 1990, Wild, Eckart, Wiklind 1997) -
the molecular gas takes over and becomes the dominant component of the
neutral (atomic and molecular) ISM.

\begin{figure}
\centerline{\includegraphics[width=28pc]{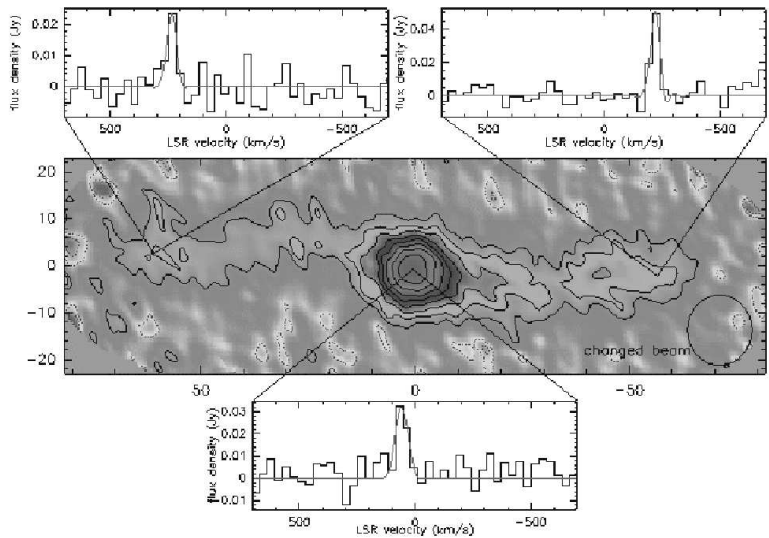}}
\caption{ A summary of recent Plateau de Bure observations om CO(1-0) line
emission towards NGC~3718.}
\end{figure}

To understand the evolution of galaxies it is essential to study the dynamics
of the molecular gas. It is believed that the formation of bulges is coupled to
gas relocation due to a barred potential.
For the fueling process of the nuclear activity like
Active Galactic Nuclei (AGNs) or nuclear starbursts, the molecular
gas has to be moved from large radii (kpc-scale) to a few parsecs.
Therefore it is important to understand the responsible transport mechanism
i.e. to investigate structure and kinematics of the atomic and molecular gas.
How gas may easily be transported towards the center of such disks
has recently been shown by Duschl et al. (2000).
The authors suggests that viscosity within such gaseous disks may
provide an efficient AGN fueling mechanism.
\\
In the course of our ongoing study of the dynamics of active galaxies NGC~3718
is a special case. Its gas dynamics on kpc scales is dominated by a strong warp
possibly due to interaction with an outer galactic halo.
Recent observations of H$_2$O masers near the central engine in
two nearby Seyfert galaxies show that the central 1 pc disks are 
warped.
Our recent result of our high angular ($\sim$ 0.7'')
interferometric data on nearby Seyfert galaxies indicate that the CO gas
in the inner 500 pc might be warped as well 
(Schinnerer, Eckart \& Tacconi 2000a,
Schinnerer et al. 2000b, Eckart \& Downes 2001).
A detailed modeling of the molecular gas as seen using the IRAM 30m telescope
and the Plateau de Bure Interferometer (Fig.2) has been carried out and is
described in detail in a forthcoming paper by
Hartwich et al. (2002) and Krips, Pott et al. (2002). \\
Here we present shortly the most important results:\\ 
({\em i}) Based on our high resolution data of the molecular gas, we could observe, that also the inner
gas disc is warped down to 20", which was unresolved by the model of
Schwarz (1985), but is consistent with the newer HI-data of Verheijen \&
Sancisi (2001), conducted with a comparable small beam size at the WSRT.\\
({\em ii}) We estimated a rotation curve, corrected for the inclination of the
  tilted gas rings, which is constant $(235\pm15 \frac{km}{s})$ between 20"
  and 100", i.e. the central region, where the molcular gas is the dominant
  partition of the neutral gas. \\
({\em iii}) Our model explains the observed {\em warp}-features of the
  integrated line-intensity map (Fig.1) and of the pv-diagram along a
  PA=-66degr convincingly as {\em orbit crowding} and transits smoothly to the
  model of the outer HI by Schwarz (1985). \\
High resolution kinematical three dimensional models of the gas distribution as the one presented in Hartwich et al. (in
prep.) can help to understand the causes for the found warp and the history of
the galaxy by analyzing the evolution of the precessing gas rings of the {\em
  tilted ring}-model.

\end{article}
\end{document}